\documentclass[11pt]{emulateapj}
\shorttitle{The RCT 1.3-meter Photometric Calibration}
\shortauthors{Strolger et al.}

\usepackage{natbib}
\usepackage{nicefrac}
\usepackage{ulem}
\usepackage{xcolor}
\renewcommand{\deg}{^{\circ}}
\begin{document}
\title{The RCT 1.3-meter Robotic Telescope: Broad-band Color Transformation and Extinction Calibration}
\author{L.-G.~Strolger\altaffilmark{1,2}, A.~M.~Gott\altaffilmark{1}, M.~Carini\altaffilmark{1}, S.~Engle\altaffilmark{3}, R.~Gelderman\altaffilmark{1}, E.~Guinan\altaffilmark{3}, C.~D.~Laney\altaffilmark{1}, C.~McGruder\altaffilmark{1}, R.~R.~Treffers\altaffilmark{4}, 
and D.~K.~Walter\altaffilmark{5}}
\email{strolger@stsci.edu}
\altaffiltext{1}{Western Kentucky University, Bowling Green, KY 42101}
\altaffiltext{2}{present address: Space Telescope Science Institute, Baltimore, MD 21218}
\altaffiltext{3}{Villanova University, Villanova, PA 19085}
\altaffiltext{4}{Starman Systems, LLC., Alamo, CA 94507}
\altaffiltext{5}{South Carolina State University, Orangeburg, SC 29117}

\begin{abstract}
The RCT 1.3-meter telescope, formerly known as the Kitt Peak National Observatory (KPNO) 50-inch telescope, has been refurbished as a fully robotic telescope, using an autonomous scheduler to take full advantage of the observing site without the requirement of a human presence. Here we detail the current configuration of the RCT, and present as a demonstration of its high-priority science goals, the broadband {\it UBVRI} photometric calibration of the optical facility. In summary, we find the linear color transformation and extinction corrections to be consistent with similar optical KPNO facilities, to within a photometric precision of 10\% (at $1\sigma$). While there were identified instrumental errors likely adding to the overall uncertainty, associated with since-resolved issues in engineering and maintenance of the robotic facility,  a preliminary verification of this calibration gave good indication that the solution is robust, perhaps to a higher precision than this initial calibration implies.  The RCT has been executing regular science operations since 2009, and is largely meeting the science requirements set in its acquisition and re-design.

\end{abstract}

\section{Introduction}
Robotic telescopes provide a fast and efficient means to carry out various observations, especially for transient phenomena or for monitoring work, while mitigating significant and often prohibitive travel costs associated with classical observing. To this end, the 1.3-meter telescope at Kitt Peak has been restored and refurbished to operate as a fully autonomous optical facility~\citep{Gelderman:fk}. The Robotically Controlled Telescope (or RCT) is operated by an automated and efficient scheduler, and equipped with a suite of broad and narrow-band filters to enable `snapshot' imaging and monitoring on a range of time-scales. The telescope supports the broad interests of the investigators and institutions comprising the RCT Consortium, and has already contributed to the locating and monitoring of gamma-ray bursts and supernovae~\citep{McGruder:2004zr, Gouravajhala:2012lr,Gott:uq},  ephemerides of minor planets (e.g., \citealt{Buie:fk}), and the monitoring of variable stars  and active galaxies~\citep{Engle:2011lr, Lefevre:2005fk, Marchenko:2004lr, Raiteri:2006qy,Carini:2004ly}. The RCT is also regularly used in undergraduate research at Consortium institutions, and frequently contributes to science education and outreach in programs such as the Arizona Astronomy Camp~\citep{McCarthy:2012}.

Precise photometry is the highest priority to essential all such science cases. Here, we detail a preliminary photometric calibration of the RCT 1.3-meter, in the Johnson/Bessell {\it UBVRI}-bands, from data obtained from 2010 March to 2012 January.  This work, in effect, serves as the first science verification for broadband photometry with the RCT. It is an automated  analysis of numerous and periodic observations of equatorial standard stars taken throughout the start of science operations. This analysis was done largely without {\it a priori} corrections or rejections of data due to short-term variations in observing conditions, as much of this information was not readily traced (e.g., via a photometric monitor), as is discussed later in this manuscript. We do, however, attempt to isolate data of the best photometric quality for this analysis, and provide a verification of the precision of the derived extinction and color transformations for the facility.   We also detail possible systematic errors and additional caveats in using these photometric corrections in other analyses. 

In all, the RCT refurbishment project has been a success, meeting nearly all of its initial science requirements. This manuscript details the current configuration of the facility, and provides an initial broadband photometric calibration. In this section we give a brief history of the facility, and the science motivations behind the refurbishment. In \S2 we provide an overview of the facility, detailing the hardware and software essential to its automation.  In \S3 we detail the broadband {\it UBVRI} calibration program, its results, and caveats. And in \S4 we provide an overall assessment summary of the refurbishment.

\subsection{A Brief History of the RCT}
The KPNO 50-inch was developed in the early 1960's by the Space Sciences Division at Kitt Peak National Observatory (KPNO) as an engineering research platform for the development of remote control protocols for future orbital telescopes. It was to be operated on Kitt Peak remotely from Tucson, AZ via commercial telephone lines, in a concept  in-line with today's robotic telescopes. The telescope was installed in its present location in 1965, and operated as a remote observatory testbed \citep{Maran:1967fk}, until 1969 when the facility was converted into a full-time on site visitor facility. The telescope was then fitted with an aluminum coated ceramic-vitreous primary mirror cell replacing a nickel-coated cast aluminum cell~\citep{Maran:1967qy} that warped with changes in ambient temperature. For the next 25 years, the KPNO 50-inch served as an optical and near-infrared facility for visiting astronomers, as well as a testbed for National Optical Astronomy Observatory (NOAO) near-infrared instruments, such as the Infrared Imager (IRIM), the Simultaneous Quad Infrared Imaging Device (SQIID), the Cryogenic Spectrograph (CRSP), and the Cryogenic Optical Bench (COB). However, budget constraints eventually forced decommissioning of the 50-inch as part of KPNO portfolio in 1996. In 2000, the RCT Consortium successfully bid to operate the telescope (dubbed the Robotically Controlled Telescope), and completed its full refurbishment as a robotic instrument in 2009. Tests of the engineering and optical quality continued through 2012, over which time science programs such as this calibration program also commenced. It is now fully operational and open for regular use by the RCT Consortium partners.

\subsection{Overview of Science Requirements for the RCT}
The refurbishment and automation of the telescope was driven by a group of researchers, most of whom from primarily undergraduate institutions, with complementary science goals  that would benefit from an automated 1-meter direct imaging facility. Broadly, the envisioned projects required precise photometric monitoring ($\sigma_{\rm mag}\la 0.1$) of relatively bright targets ($\la20$ mag), in visits of a few minutes often executed periodically over long timespans (weeks to years). The details of each project set different requirements for photometric precision, often weighed against acquisition constraints and practical limits on the overall versatility of the automatic scheduling. Here we focus on a few projects which demonstrate some of these challenges, and thus exemplify the science requirements for the RCT.

\noindent {\bf Measuring Transiting Extrasoloar Planets}: An important goal in probing stellar system evolution is to at least detect large planets about relatively small stars, as for example Neptune-sized planets about main sequence F or G stars. Through the  transiting method, the change in apparent brightness, or transit depth ($\Delta m$),  is at best dependent on the relative cross sections of each component by:
\begin{equation}
	\Delta m = -2.5\,\log\biggl(1-\frac{R_p^2}{R_\star^2}\biggr),
\end{equation}
where $R_p$ and $R_\star$ are the radii of the planet and star respectively. A Neptune-like planet transiting an F-star would result in a $\Delta m$ of, at most, $\sim 0.001$ mag. This objective places strong constraints on photometric precision for the telescope.  Single-exposure estimates based on the RCT exposure time calculator (RCT-ETC) suggests $\sigma_{\rm mag}\simeq0.001$ should be achievable for bright (roughly $V\la14$ mag) targets in 2 to 3 minute exposures~\citep{McGruder:2004zr}. The RCT-ETC ({\tt http://rct.wku.edu/kittpeak/etc.html}) is based on the IRAF task {\tt ccdtime}, and provides approximate exposure times, S/N, and magnitude limits based on optimized estimates of the site and telescope conditions. We discuss comparisons to real data in \S\ref{sec:etc}. 

\noindent {\bf Ephemerides of Minor Planets}: Refining the ephemerides of minor objects in the solar system system requires precise astrometry, which is often complicated by the non-sidereal motion of the object itself. The goal is to obtain sufficient signal-to-noise in exposures short enough to keep star field trailing significantly below the FWHM of the point spread function (PSF). For many comets and asteroids this limits exposure times to only a few tens of seconds, which should be sufficient for $\sigma_{\rm mag}\le0.1$ on objects as faint as $V\sim17$~mag.

\noindent {\bf  Monitoring Active Nuclei}: Blazar variability monitoring, particularly on short timescales, can provide important clues to the nature of the supermassive black holes which they harbor.  Well-sampled light curves of the variation in optical continuum provide probes the size of emission regions that, when combined with complementary data from gamma-ray to radio wavelengths, provide strong constraints for model conditions of these AGN. These variations, however,  are unpredictable and  irregular occurrences, lasting several minutes or several years, and thus requiring a level of periodic monitoring which often challenges classical scheduling at ground-based telescopes. The objective for the RCT is keep to an observing cadence for some programs, and use other periodic programs as fillers to keep the overall schedule efficient. The automated scheduling of the RCT, as discussed further in \S\ref{sec:insgen}, is efficient enough to wedge small programs between optimal windows of larger ones, repurpose windows of failed observations, and utilize fractions of nights in marginal or sub-optimal conditions which might otherwise be lost. This, in combination with options for repeated requests allows a regular monitoring of blazars {\it  ad perpetuum}, and provides well-sampled long-term light curves for complementary studies with, for example, RXTE, VERITAS, and Fermi~\citep{Carini:2004ly}. 

\noindent {\bf Light Curves of Supernovae and Gamma-Ray Bursts}: Both gamma-ray bursts and supernovae are now routinely discovered at a rate several hundred per year, respectively, now enabling a more comprehensive understanding of the nature of these extremely energetic explosions. There is still much to ascertain on the most and the least energetic ranges of both supernovae and gamma-ray burst optical transients, especially at early times, or within hours of explosion: most critically for gamma-ray burst of the shortest duration. Doing so for a sufficiently large sample will likely require coordinated action from rapid-response observatories, ideally with queues that can be appended to or reconstructed automatically (within minutes) to catch high priority events~\citep{McGruder:2004zr}.  As discussed further in \S\ref{sec:insgen}, the RCT scheduler is designed to take relative priority of observations into account when building its schedule, which is rebuilt frequently to accommodate new targets throughout night. Coordination on targets of opportunity will eventually come from large scale ``all-sky'' surveys, such as PanSTARRS, SNfactory, and Palomar Transient Factory (and eventually LSST), which will act as ``event brokers'' for the community via VOEvents, GCN, ATEL, CBET, or other electronic announcements~\citep{hwtu2}.

\noindent{\bf Narrow-Band Imaging of Resolved Nebulae and Comets}: The RCT is equipped with two large filter wheels, allowing in addition to standard broadband filters, immediate access to one of two sets of narrow-band interference filters. The narrow-band filters are centered on diagnostic nebular emission lines from H$^+$, He$^{++}$, S$^+$, N$^+$, and O$^{++}$ in one set, and molecular diagnostic lines (with complementary continuum) useful for comets and minor planets in a second set. These narrow-band filters extend the versatility of the RCT, allowing 2-dimensional, spacially resolved maps of extinction (H$\alpha$/H$\beta$), electron temperatures and densities, and ionization maps  for nearby nebulae, and extensive coverage of comets pre- and post-perihelion to monitor outburst and fragmentation, determine volatile production rates, and trace the evolution of the comae~\citep{Walter:2004fr}.

\section{The Current Configuration of the RCT}
The RCT 1.3-meter is the original Boller \& Chivens Schmidt-Cassegrain on a German equatorial mount, with the aluminum coated CerVit cell fitted in 1969. The facility retains the original dome, and still uses the same mechanical system in the dome drive and telescope drives, although most motors have been replaced with newer models. Most modifications are to the control system to provide a higher-level control and management of the telescope, enabling its use as a automated facility.  In this section we detail this current design of the RCT. For convenience, we provide a brief overview of the facility in Table~\ref{tab:props}.

\begin{table*}[ht]
\begin{center}
\caption{\sc RCT facility Overview\label{tab:props}}
\vspace{0.1in}
\begin{tabular}{ll}
\tableline
\tableline
Location & Kitt Peak National Observatory, Arizona\\
&$31\deg57\arcmin12\arcsec$ N, $111\deg34\arcmin$ W (WGS84)\\
Elevation& 2040 meters above sea level (approx. 800 meters above horizon)\\
Optics & 1.3-meter Schmidt-Cassegrain, $f/13.5$ secondary\\
Plate Scale & $11\arcmin.89$ mm$^{-1}$, $0\arcsec.285$ pixel$^{-1}$, $9\arcmin.6\times9\arcmin.6$ FOV\\
CCD& SITe 2048 x 2048 CCD, 24~$\mu$m pixels\\
& 4 available amplifiers, 1 currently in use (120 sec read time)\\
&Gain (amp C) = 2.56 e$^-$ DN$^{-1}$, Read Noise (amp C) = 15.05 $e^-$\\
&Minimum Full Well signal $\approx150,000$ $e^-$\\ 
& dark current $\approx 1\times10^{-5}\,\, e^-$ s$^{-1}$ pixel$^{-1}$\\
Filters& 5 broadband, and 21 narrow-band\\
&Two active wheels with 16 total positions continuously available\\
Modes of Operation& Robotic queue scheduled; \\
&Remote or on-site, scripted or standard observing\\
\tableline
\end{tabular}
\end{center}
\end{table*}

The newest aspect of the RCT 1.3-meter is its telescope control system (TCS) and network of automated monitors 
developed by Starman Systems, LLC. The TCS is designed around an open industrial panel, using a 16-axis Delta Tau Turbo PMAC to control motors and encoders, and an ADAM I/O module for additional digital and analog functions, and PLDs for complex relay logic problems such as the mirror covering opening sequence and dome slit control. Most of  the low level automation software is written in {\tt C++},  developed at U.~C.~Berkeley for the Katzman Automatic Imaging Telescope at Lick Observatory~\citep{Richmond:1993fk}. Basic functionality uses ASCII protocol communicating over Berkeley sockets. 

The automated system coordinates with a few network services to improve its accuracy. Time data is collected using the Network Time Protocol daemon, and celestial calculations are made from in-house code and  checked against SOFA routines.\footnote{\footnotesize Software Routines from the International Astronomical Union Standards of Fundamental Astronomy Collection. ({\tt http://www.iausofa.org})} JPL Horizons~\citep{Giorgini:rm} HTTP calls\footnote{\footnotesize {\tt http://ssd.jpl.nasa.gov/horizons.cgi}} are used for ephemerides of solar system objects and to derive non-sidereal track rates. The {\tt TPoint} package\footnote{\footnotesize {\tt http://www.tpsoft.demon.co.uk}} is used for all pointing modeling. All other software is in-house and non-proprietary. 

The automated system also coordinates with a few on-site monitors. Optional guiding is provided by an automated guider camera, which locates bright stars within a 2$\deg$ region of the pointing using the Yale Bright Star Catalog,\footnote{\footnotesize{\tt http://tdc-www.harvard.edu/catalogs/bsc5.html}} zeros pointing to within $20\arcsec$ of the original pointing, and relays continuous corrections to the telescope drive rates. The system also receives real-time, on-site meteorological information from a Vaisala automatic weather station,\footnote{\footnotesize{\tt http://www.vaisala.com}} which relays humidity, wind, and rain information to safety protocols in the dome slit controller. 

All image metadata are archived in a SQL database on-site, which also manages observation requests for the automated scheduler, described below.  Raw images are stored in the Starbase Database, a managed repository at Western Kentucky University available to partner institutions.

\subsection{Optical Camera and Filters}
The RCT camera operates with a $2048\times2048$ SITe SI-424A backside-illuminated CCD with 24~$\mu$m pixels. The CCD is housed in an IRLabs dewar,\footnote{\footnotesize{\tt http://www.infraredlaboratories.com}} which is cooled to $172$K via a closed-cycle, PT-14 gas Polycold refrigeration system from Brooks Automation, Inc.\footnote{\footnotesize{\tt http://www.brooks.com}} The camera sits at the $f/13.5$ Cassegrain focus,  to produce a plate scale of 11.\arcsec89 ~mm$^{-1}$ (0.\arcsec285 ~pixel$^{-1}$), and an unvignetted $9.\arcmin6\times9.\arcmin6$ field of view. 

The filter-wheel assembly sits above the dewar window, housing a pair of 9-slot filter wheels, within each are 8 circular mounts for 4.5-inch diameter filters and one unthreaded open mount, to provide a total of 16 possible filter options (and one double-open position) continuously available. Table~\ref{tab:filters} provides a complete list of the filters available, along with effective wavelengths, bandwidths (or FWHM), and maximal throughputs.

\begin{deluxetable*}{lcccl}
\tablecolumns{4}
\tablewidth{0pt}
\tablecaption{\sc RCT Filter Set}
\tablehead{\colhead{Filter}&\colhead{Effective $\lambda$ (\AA)}& \colhead{Bandwidth (\AA)}&\colhead{Max. Trans. (\%)} &\colhead{Model \#}\label{tab:filters}}
\startdata
\sidehead{{\it Broadband:}}
$U$\dotfill&3573&900& 75&ANDV8303\\
$B$\dotfill&4265&931& 66&ANDV8666\\
$V$\dotfill&5434&751& 89&ANDV8667\\
$R$\dotfill&6487&1016& 84&ANDV8668\\
$I$\dotfill&8129&1728& 96&ANDV8669\\
\sidehead{{\it Nebular Narrowband:}~\tablenotemark{a}}
HeII\dotfill&4689&33&66&ANDV8163\\
Green Continuum \#1\dotfill &4807&53&62&ANDV8164\\
H$\beta$\dotfill&4864&35&63&ANDV8165\\
$[$OIII$]$\dotfill&5009&25&62&ANDV8166\\
Green Continuum \#2\dotfill &5313&96&70&ANDV8167\\
$[$NII$]$ \#1\dotfill &5760&30&60&ANDV8168\\
Red Continuum\dotfill &6459&83&63&ANDV8169\\
H$\alpha$ Narrow\dotfill&6565&14&38&ANDV8281\\
H$\alpha$ Wide\dotfill&6561&25&51&ANDV8282\\
$[$NII$]$ \#2\dotfill &6594&29&42&ANDV8283\\
$[$SII$]$ \#1\dotfill &6716&13&51&ANDV8284\\
$[$SII$]$ \#2\dotfill &6729&14&55&ANDV8285\\
\sidehead{{\it Comet Narrowband:}}
OH\dotfill&3090&62&49&Barr Lot 4108\\
UV Continuum\dotfill& 3448&84&72&Barr Lot 4108\\
CN\dotfill&3870&62&80&Barr Lot 1109\\
C$_3$\dotfill&4062&62&69&Barr Lot 4108\\
Blue Continuum\dotfill&4450&67&76&Barr Lot 4108\\
C$_2$\dotfill&5141&118&82&Barr Lot 4108\\
Green Continuum \#3\dotfill&5260&56&80&Barr Lot 1009\\
NH$_2$ Continuum\dotfill& 5660& 12&67&Barr Lot 1109\\
NH$_2$\dotfill&5721&85&80&Barr Lot 4408\\
\enddata
\tablenotetext{a}{\footnotesize The nebular filter set was re-scanned in the laboratory in 2012 March.}
\end{deluxetable*}

\begin{figure*}[t]
	\centering\includegraphics[width=5.5in]{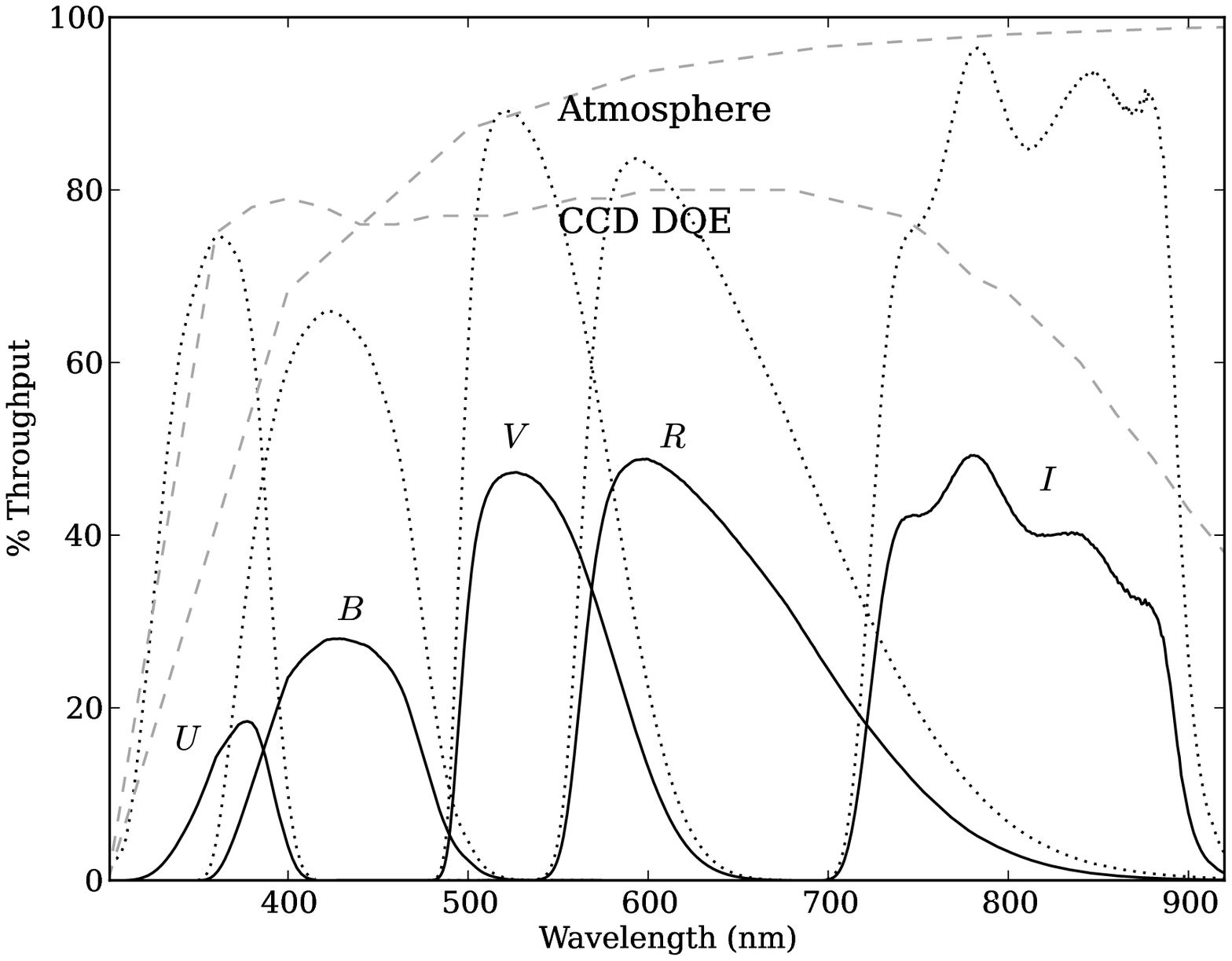}
	\caption{\footnotesize The nominal RCT Johnson/Bessell {\it UBVRI} filter responses. The dotted lines show the laboratory transmissions for each filter, and the solid lines show the throughputs after multiplying by the CCD quantum efficiency (at $T=172K$) and nominal atmospheric transmission at zenith (dashed lines). The throughputs are also corrected by measures of the primary mirror reflectance (approximately $87\%$ from 4000~{\AA} to 7000~{\AA}) and dewar window transmission (approximately $95\%$ from 4000~{\AA} to $1\mu$m).\label{fig:filter}}
\end{figure*}

The RCT broadband set consists of  Johnson/Bessell {\it UBVRI} filters from Andover Corporation.\footnote{\footnotesize{\tt http://www.andovercorp.com}} Figure~\ref{fig:filter} shows the nominal and effective filter throughputs, corrected for the CCD quantum efficiency, nominal atmospheric transmission at zenith from a smoothed {\tt MODTRAN5} model~\citep{Berk:2006lr}, dewar window transmission, and primary mirror reflectance, as measured in 2009.  These are ideal measures, as factors associated with degradation and frequency of maintenance reduce the overall system response, as discussed further in \S~\ref{sec:etc}.

\subsection{Scheduled Observations via the Autonomous Scheduling Algorithm}\label{sec:insgen}
While the RCT 1.3-meter can be operated on-site or remotely, its main function is in a full-robotic imaging mode. The autonomous scheduling algorithm is an essential component of this observatory, and has been designed to manage numerous requests with various observing constraints. It is built from the {\tt INSGEN} list generator and process spawner originally developed for the Berkeley Automatic Imaging Telescope~\citep{Richmond:1993fk}, which makes scheduling decisions through a least-cost insertion algorithm.

The method of the {\tt INSGEN} scheduling algorithm (or scheduler), as fully described in \cite{Richmond:1993fk}, is as follows. Users provide ``requests'' which detail observation information such as the target(s), the number of exposures, the exposure times, and filters, and the observing constraints such as the range of airmass, minimum angle of Sun/Moon avoidance, or specific start or end times. Users also assign a priority to their requests, which in principle can be managed by partner fractions. The scheduler calculates ``windows of opportunity'' for each target in each request based on the target's rise and set times, the block length of the request, readout and slew times, sunrise and sunset, and other constraints. 

A ``best time'' is defined within each window, which is generally centered at the meridian transit unless otherwise constrained. The scheduler sorts the request list by priority, and  reserves blocks at best times according to priority. As the scheduler proceeds through the prioritized list, it searches outward from a given request's best time for free blocks within the target's window of opportunity. If no free window can be identified,   the scheduler then attempts to accommodate the block by pushing higher priority requests away from their optimal spots without pushing them out of their observable windows or reordering the sorted list. If this cannot be accomplished, the request with the lower priority is skipped until the following scheduling instance, in which it discards the original list and starts again.

Scheduling instances are implemented periodically throughout the daytime hours, and each time an observation has completed. With each instance, the previous observation is rejected from the list (if applicable), and the allowable windows are re-constrained by either sunset or the current time, whichever is latest. This allows for real-time ingesting of new observation requests at any time, even while the telescope is observing, and significant flexibility to keep a highly efficient schedule, should a previous request have shorter execution window than anticipated (usually slew times shorter than the estimated 2 minute intervals), or fail due to lack of available guide stars. Scheduling instances take little time to execute, usually less than 10 seconds for the hundreds of active requests currently in the queue.

\begin{figure*}[h]
	\centering\includegraphics[width=5.5in]{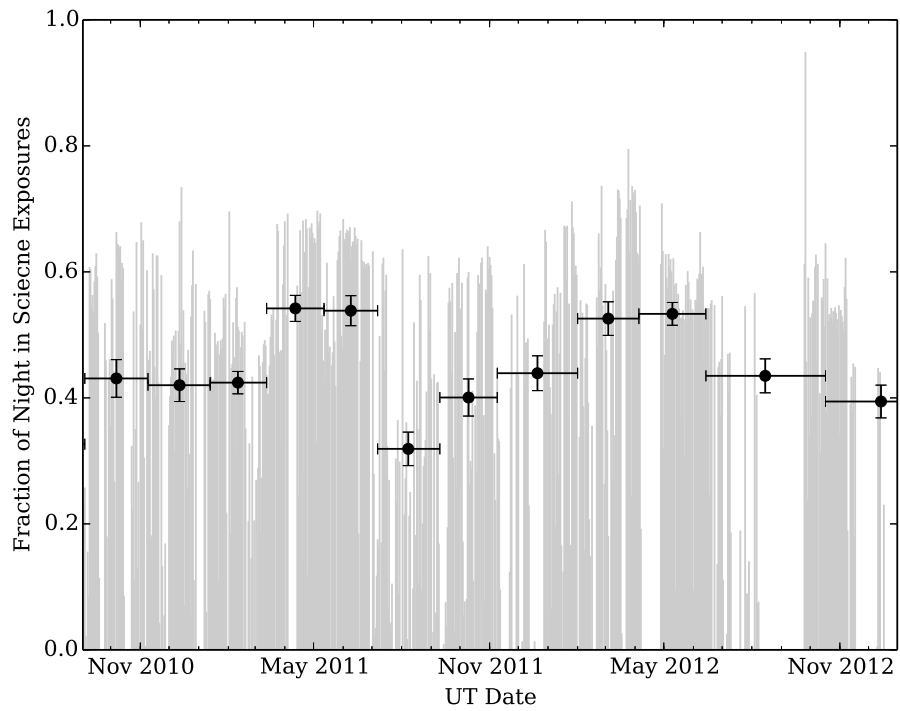}
	\caption{\footnotesize An illustration of the scheduling efficiency of the RCT for a selected date range. The gray impulses indicate the fraction of each night the telescope exposed on targets, determined from the sum of exposure times and read-out times divided by the length of night between evening and morning astronomical ($18\deg$) twilights. Field acquisition, time lost to weather, and instrument down times account for the remainder of these fractions. The black points are binned averages running every 50 nights in which data is taken. ~\label{fig:eff}}
\end{figure*}
The {\tt INSGEN} scheduler operates the RCT with highly efficient observing plans, over periods of continuous observing, or without instrument outages or poor weather. As a measure of the telescope's efficiency, we show in Figure~\ref{fig:eff} the fraction of each night the imager exposed on fields, which was generally between 40\% and 60\%. This efficiency figure includes the CCD readout time for each exposure, as this is set by the binning mode or similar details for a specific observation. But the fraction omits the slew time to each field, filter setup, or guide star acquisition (when applicable) as these are limitations more inherent to the facility. The remainder of the time was lost to weather and instrument servicing, with less than 10\% spent on field acquisition and other overheads.

The observation requests are collected and logged in an on-site database, populated by users via a web interface available to partners. There are several convenience options at the interface, including target  resolution via NED, SIMBAD, JPL Horizons, or via the expanding RCT database.  Users can monitor the schedule via the same web interface, seeing both tabular and graphical representations of the scheduled, skipped, and executed requests.

\section{The Photometric Calibration of RCT {\it UBVRI} Filters}
The purpose of this paper is to not only give an introduction to the RCT, but to provide measure of its photometric quality. Observations of  \cite{Landolt:2009lr} standard stars were obtained in several sequences in from 2010 May 25 to 2012 Dec 31. This program took advantage of the scheduler efficiency to queue periodic observations of targets at the scheduler's lowest priority, thereby executing when the scheduler filled gaps between ideal meridian crossings of higher priority programs. The sequence executed 3 successive images in each passband, with exposure times of 30~s, 20~s, 10~s, 10~s, and 10~s, in each {\it U, B, V, R,} and {\it I}, respectively. The exposures were not guided to reduce overheads.

The sequence was taken without observing constraint, allowing for a modest range of airmass and cloud conditions for broad assessment.  However, as the goal of the scheduler is to observe targets as close to meridian whenever possible, very few observations were actually executed at large [$\sec(z) > 1.5$] airmass. The supernova in M101 in 2011 August (SN~2011fe; \citealt{Nugent:fk}) presented an important target of opportunity, and its low altitude at KPNO necessitated a set of high airmass standards for accurate calibration of its observations.\footnote{\footnotesize The RCT 1.3-meter {\it UBVRI} photometry of SN~2011fe is presented in~\cite{Gouravajhala:2012lr} and \cite{Gott:uq}.} An additional standard sequence at airmass $> 2.2$  was therefore added to the request over the observable duration of SN~2011fe.

\begin{figure*}[h]
	\centering\includegraphics[width=5.5in]{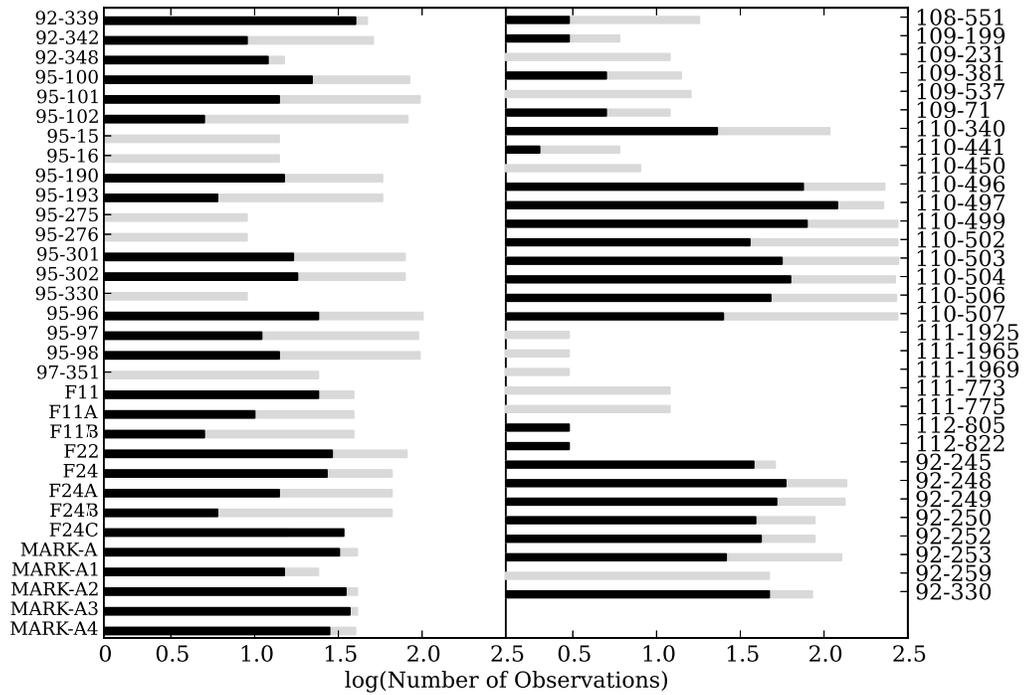}
	\caption{\footnotesize Total number of \cite{Landolt:2009lr} standard star observations (gray bars) in the $V$-band, for each standard indicated on the ordinate axis. Black bars indicate the number of observations ultimately used in the final analysis.\label{fig:freq}}
\end{figure*}

As with most observations at the RCT, these standard star sequences  were taken with the CCD binned $2\times2$, to take advantage of the oversampling of the typical PSF (approximately 1.\arcsec5 to 2.\arcsec5~FHWM) at the site. This results in a pixel scale of $0.\arcsec571\,\pm\, 0.\arcsec001$ pixel$^{-1}$, verified against catalog solutions in a process described further in \S\ref{sec:phot}. Figure~\ref{fig:freq} indicates which~\cite{Landolt:2009lr} stars were observed and how often. Ultimately, $5,600+$ {\it UBVRI} sequences of standard stars were taken,  of which approximately $1,000 - 2,000$ stars per filter were used in the final analysis.

\subsection{Data Reduction, Photometry, and Calibration}\label{sec:phot}
The RCT executes a calibration sequence in a separate request, taken nightly or as priority allows. Bias images were taken in a dark dome (after twilight) as a precaution, although the instrument shows no evidence of light leaks and the dark current is estimated to be less than $1\times10^{-5}$ e$^{-}$ s$^{-1}$. Flat-field correction images were derived from images of a lamp-illuminated screen on the interior of the dome, which are done as a matter of course in at least each of the broadband filters. All science data in this analysis were bias and dome flat-field corrected using the IRAF {\tt imred.ccdred} package, with respective nightly correction images. Comparisons of a few dome-flat and bias images yielded a nominal read noise of 15.1 e$^-$, and gain of 2.6 e$^-$ DN$^{-1}$. 

Raw RCT images have only an approximate World Coordinate System map to their pixels. To accurately correlate sources with known positions of standards, we pattern matched detections (via {\tt SExtrator}, \citealt{Bertin:yg}) with the USNO B1 catalog~\citep{Monet:eu}, using a triangle-matching algorithm which solves for linear source-to-source transformations, including shifts, rotations, and changes in scale, in which the plate scale is a free-parameter. This automated fitting required greater than approximately five detections per image, evenly distributed, for accurate matching (to RMS $<0.\arcsec2$). For most observations in the redder passbands of these standard fields, where there is greater sensitivity and a larger number of red sources, this criteria was easily met. But for bluer passbands where there were fewer detected sources, the failure rate in direct pattern matching was as much as two times more frequent than for the redder passbands, resulting in about half as many correlated measures. 

We may have had more success in pattern matching if we reduced the free-parameters in the matching algorithm, fixing the plate scale to set value for example, although it was important to independently verify the plate scale of the image plane.  We might also have made use of the back-to-back execution of the observing sequence to create ``detection frames'' from combinations of images, in the same filter or in all-filter combinations. Although that would have required some slight image-to-image forced alignment, through cross-correlation or human source identification, as small ($\sim0.\arcsec5$) drifts between exposures occur in the sequence when tracking was momentarily resets. However, the final number of matched targets were more than sufficient to precisely determine the photometric transformations, even after the rejection pass, and so it was not deemed necessary to put effort into recovering these acceptable losses.

\begin{figure*}[h]
	\centering\includegraphics[width=5.5in]{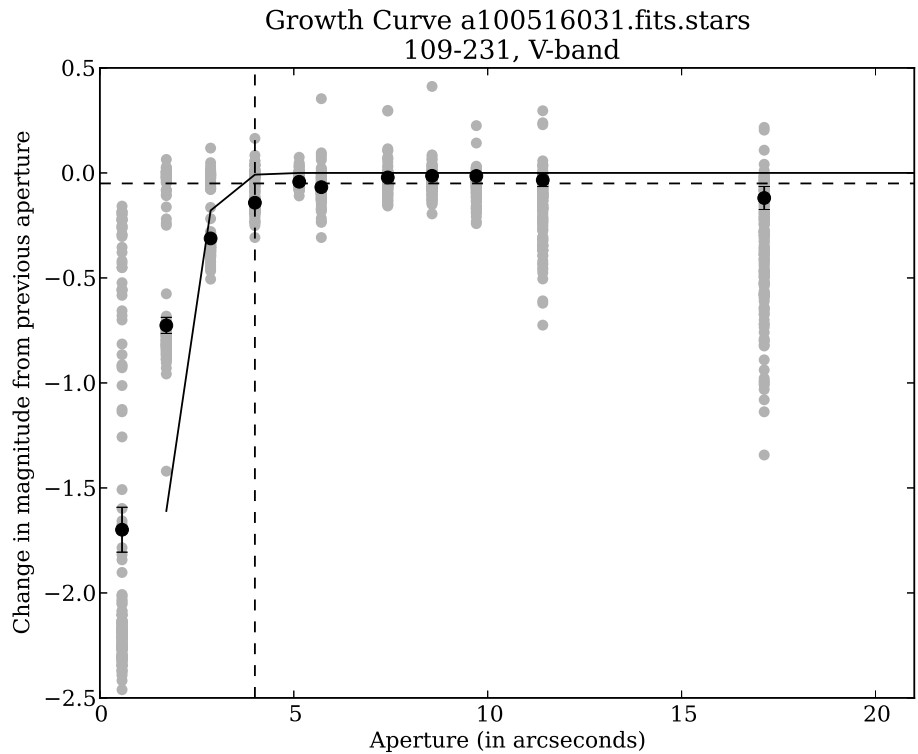}
	\caption{\footnotesize Example growth curve for stars in one {\it V}-band exposure, centered on SA~109-231. Gray points show the change in magnitude at the set aperture indicated, from the measure in the next smallest aperture, for each of the 167 detections (to $3\sigma$) in the image. Black points are the mean for all stars in each aperture. The solid line is the expected magnitude change a 2-D Gaussian with an equivalent FWHM. Dashed lines show the selected confusion aperture (abscissa), and its corresponding aperture correction (ordinate).\label{fig:growth}}
\end{figure*}

Instrumental aperture photometry was performed in a 6.5 pixel ($3.\arcsec71$) radius on rectified images, scaled by their exposure times to units of DN~s$^{-1}$. The aperture was chosen by determining the growth curve of several hundred stars in selected images of each passband, as illustrated in Figure~\ref{fig:growth}, and selecting an aperture consistently near the confusion limit ($6-7$ pixel radii, on average). This aperture was large enough to accommodate changes in the PSF due to seeing and other instrumental effects. Aperture corrections were estimated from ratios of the encircled energy of a 2-D Gaussian with symmetric FWHM equal to the mean measured FWHM of the image, at the confusion aperture relative to an infinite aperture. In all passbands, these corrections were found to be less than 0.01 mag. However, it is important to note, as can be seen in Figure~\ref{fig:growth}, that the shape of the PSF was typically more complex than the 2-D Gaussian fit. An extreme case due to instrumental failure is also shown in \S\ref{sec:caveats}. 

\begin{figure*}[t]
	\centering\includegraphics[width=5.5in]{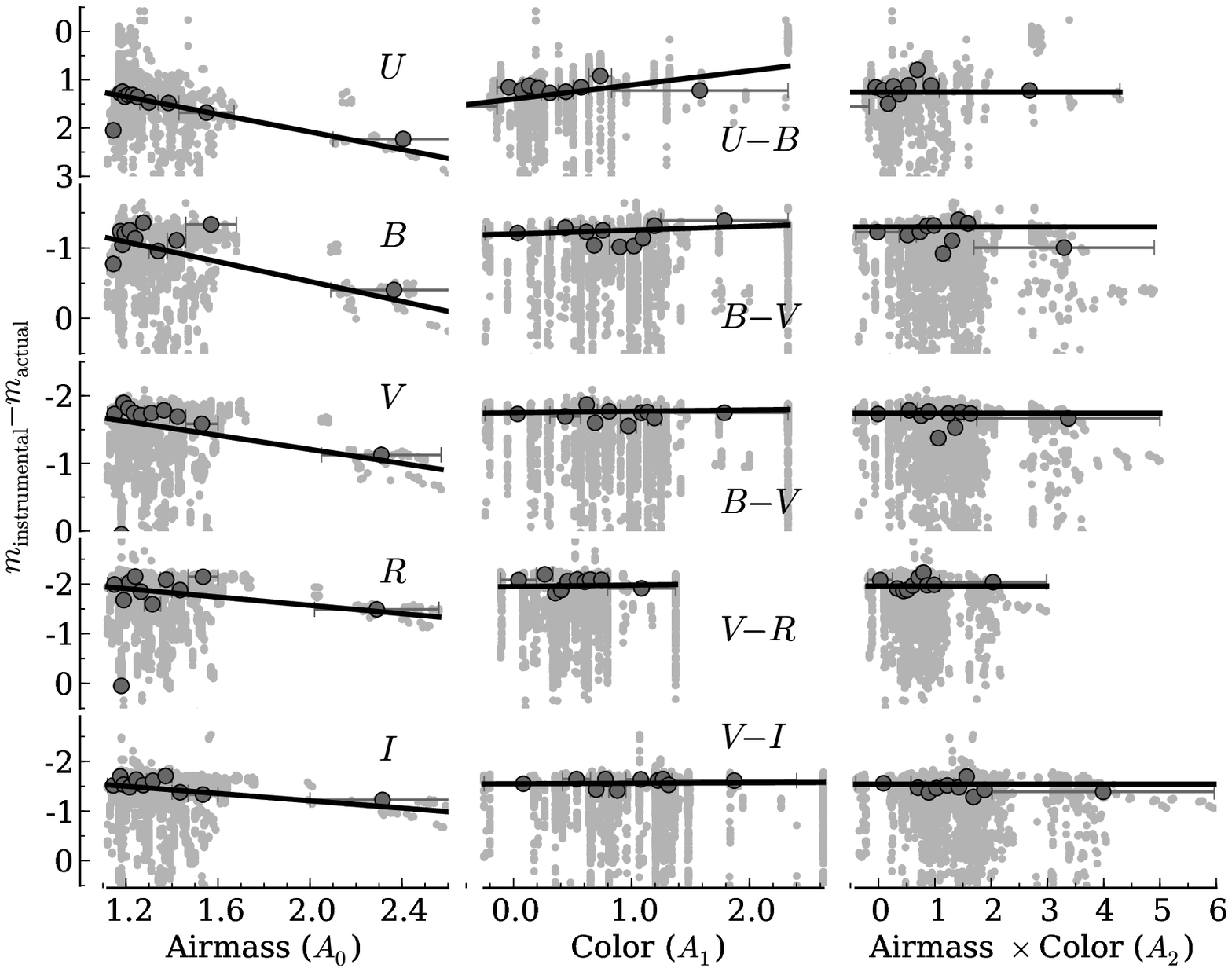}
	\caption{\footnotesize Difference in instrumental and cataloged standard star measures, as functions of fit parameters Airmass, Color, and Airmass $\times$ Color (gray points). The black points are equilibrium binned data, with approximately 250 points in each bin, where the errorbar width represents the bin width. The lines slopes are the $A_0$ and $A_1$ coefficients for each passband and color indicated. The  cross extinction-color coefficient is fixed to $A_2\equiv0.0$.}
	\label{fig:coeffs}
\end{figure*}

A first-pass calibrated magnitude ($m$) was determined in each filter from the instrumental magnitude using,
\begin{equation}
m = -2.5\log_{10}({\rm DN\,\,s}^{-1})+Z+A_0X+ A_1({\rm color})+A_2X({\rm color}),\label{eqn:phot}
\end{equation}
where $Z$ is the zero point (magnitude at 1 DN s$^{-1}$), $A_0$ is the airmass term coefficient, $A_1$ is the color term coefficient (using colors later indexed in Table~\ref{tab:coeffs2}), and $A_2$ is the airmass $\times$ color term coefficient with the same color indices. The calibration coefficients in Equation~\ref{eqn:phot} ($Z$, $A_0$, and $A_1$ respectively) were fit via the weighted Levenberg-Marquardt  non-linear least  squares  algorithm provided by the IRAF {\tt apphot} package. The cross extinction-color corrections (or $A_2$ terms) are often difficult constrain when simultaneously fit with the color transformations and first order extinctions, and leaving these terms free resulted in degeneracies that were inconsistent with the data.  It was expected, however, that these terms would be negligible, with $|A_2| < 0.02$. We, therefore, fixed at $A_2\equiv0.0$ for all filters throughout this analysis. Figure~\ref{fig:coeffs} shows the difference in instrumental and cataloged magnitudes as functions of each of the fit parameters. Equilibrium binned data (11 bins of approximately 250 measures in each bin) and the slopes of the preliminary photometric calibration coefficients are shown for comparison. Note that the fits are to the data, and not to the binned points.

\begin{figure*}[t]
	\centering\includegraphics[width=5.5in]{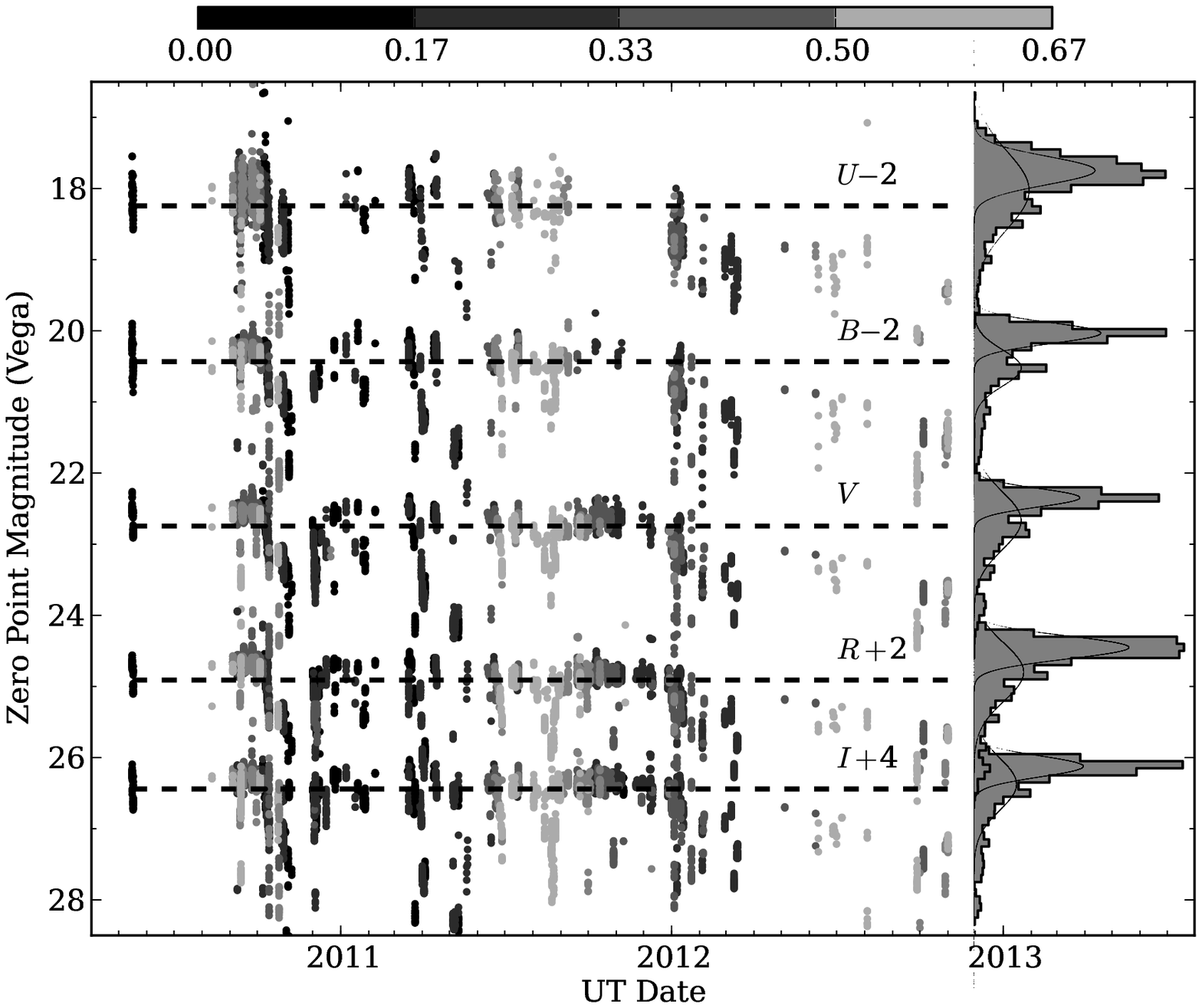}
	\caption{\footnotesize Residuals of the photometric calibration shown at the respective zero point magnitudes for each passband (dashed line). Data are shown in grayscale illustrating their relative $cloud$ value, indicated by the color bar above the diagram. Insets to the right of the figure show histograms of calibrated data (shaded), centered at the respective zero point lines. Over-plotted lines show the bi-modal Gaussian PDFs fit to the histograms. Data is offset by additional values annotated, for clarity.}
	\label{fig:zp}
\end{figure*}

Figure~\ref{fig:zp} shows the residuals of the fits, biased to the initial zero point magnitudes for each filter, with additional offsets for clarity. The RCT is equipped with a modest ``cloud monitor'', a thermoelectric infrared sky temperature sensor which compares sky temperature to ground temperatures to, in principle, derive an approximate fraction of the sky covered in clouds. The residuals in Figure~\ref{fig:zp} are shaded in grayscale by these relative $cloud$ values. As non-photometric data is inherently fainter, their measures stream downward on this  plot. It is apparent from the figure that the $cloud$ values do little to discern non-photometric (or otherwise faint) data. At present, it is unclear if the cloud monitor is too imprecise, or inaccurate due to failure, but it would seem that it is not yet a useful indicator of photometric quality, and thus insufficient for isolating photometric data for this analysis. 

\begin{figure*}[t]
	\centering\includegraphics[width=5.5in]{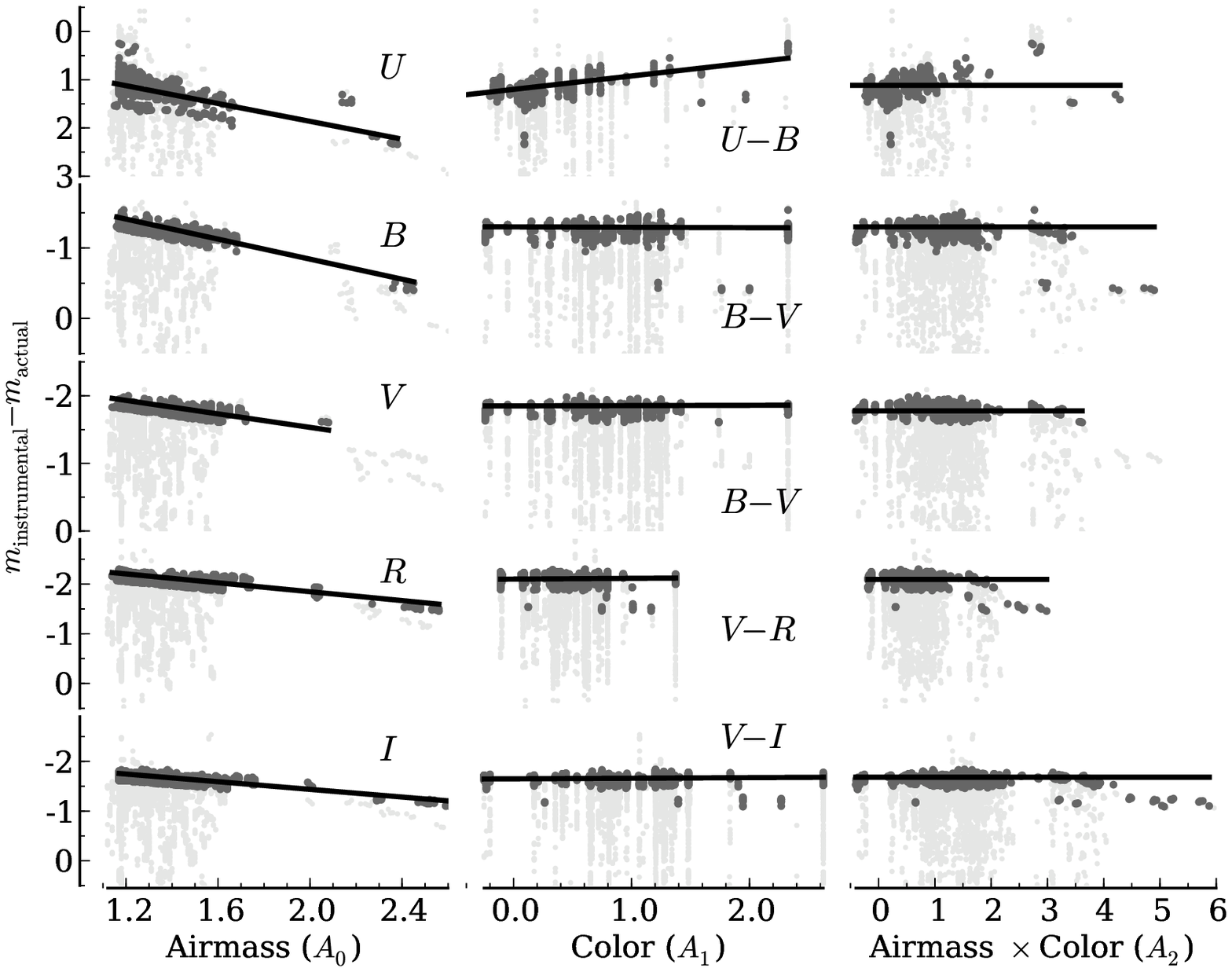}
	\caption{\footnotesize Similar to Figure~\ref{fig:coeffs}, with lightly shaded points indicating rejected data, and solid lines showing the final fit. The data shown here are not binned.}
	\label{fig:coeffs2}
\end{figure*}

The distribution of residuals in the fits do, however, provide a guide for isolating photometric data. The histograms inset to the right of Figure~\ref{fig:zp} show the distribution of the residuals relative to the zero points derived in this first pass analysis. As one would expect, these distributions grossly show modes near the  bright end of each calibrated set, with skewed smaller populations trailing toward the faint end from gradually less photometric observations. However, there also appears to be bimodality in the distributions, with fainter secondary loci located the tail of the distributions, most visibly in the $B$-band set. Such distributions could result from data obtained in systematically similar (and unfavorable) weather conditions, or from a systematic error in the functionality of the telescope. We discuss some likely possibilities, with focus on the latter, in \S\ref{sec:caveats}. For the sake of furthering this analysis, we assume these two populations are distinct, and can be reasonably separated into photometric and non-photometric (or otherwise obscured) sets. We fit bi-modal Gaussian probability distribution functions (PDFs) to the histograms, solving for the mean and standard deviation of each mode. These distributions are over-plotted on the histograms of Figure~\ref{fig:zp}.

We then performed a second-pass analysis, following the same {\tt apphot} procedure to determine calibration coefficients outlined at the beginning of this section, on only the data constrained within $1\sigma$ of the bright modes in the histograms of Figure~\ref{fig:zp}. The results of this final pass are shown in Figure~\ref{fig:coeffs2}, and and the final calibration coefficients are shown in Table~\ref{tab:coeffs2}.

\begin{deluxetable*}{lccccc}
\tablecolumns{6}
\tablewidth{0pt}
\tablecaption{\sc {\it UBVRI} Extinction and Color Transformation for the RCT~1.3-meter}
\tablehead{\colhead{Filter}&\colhead{Zero Point}&\colhead{Airmass\tablenotemark{a}}&\colhead{Color Term\tablenotemark{a}}&\colhead{$1\sigma$ Error}&\colhead{N meas.}\\
\colhead{}&\colhead{$(Z)$}&\colhead{$(A_0)$}&\colhead{$[A_1({\rm color})]$}&\colhead{}&\colhead{}\label{tab:coeffs2}}
\startdata
$U$\dotfill& 20.316 $\pm\,0.003$ & -0.924 & 0.277 $(U-B)$ & 0.10 & 910\\
$B$\dotfill & 22.550 $\pm\,0.002$ & -0.712 & -0.005 $(B-V)$  &  0.06 & 1281\\
$V$\dotfill & 22.785 $\pm\,0.002$ & -0.499  & 0.005 $(B-V)$  &  0.06& 1284\\
$R$\dotfill& 23.001 $\pm\,0.002$  & -0.448 & 0.016 $(V-R)$  &  0.07& 1993\\
$I$\dotfill& 22.513 $\pm\,0.001$ & -0.386 & 0.012 $(V-I)$  &  0.06& 2167\\
\enddata
\tablenotetext{a}{\footnotesize Standard Deviation of the Mean in Airmass and Color terms are all less than $0.001$.}
\end{deluxetable*}

\subsection{Verification of the Photometric Calibration}~\label{sec:etc}
The final derived color transformation coefficients are in excellent agreement with available data from comparable telescopes with optical imagers and similar broadband filters at KPNO, specifically the WIYN 0.9-meter and the KPNO 2.1-meter~\citep{massey:2002}. The extinction coefficients are reasonably similar, although our analysis has been carried out to larger airmasses than presented by~\cite{massey:2002}. As discussed previously, no attempt was made to fit the airmass $\times$ color terms for possible atmospheric reddening, as it was assumed the effect was negligible. The lack in notable trend in the data of Figures~\ref{fig:coeffs} and~\ref{fig:coeffs2} generally confirm this assumption.

\begin{figure*}[t]
	\centering\includegraphics[width=5.5in]{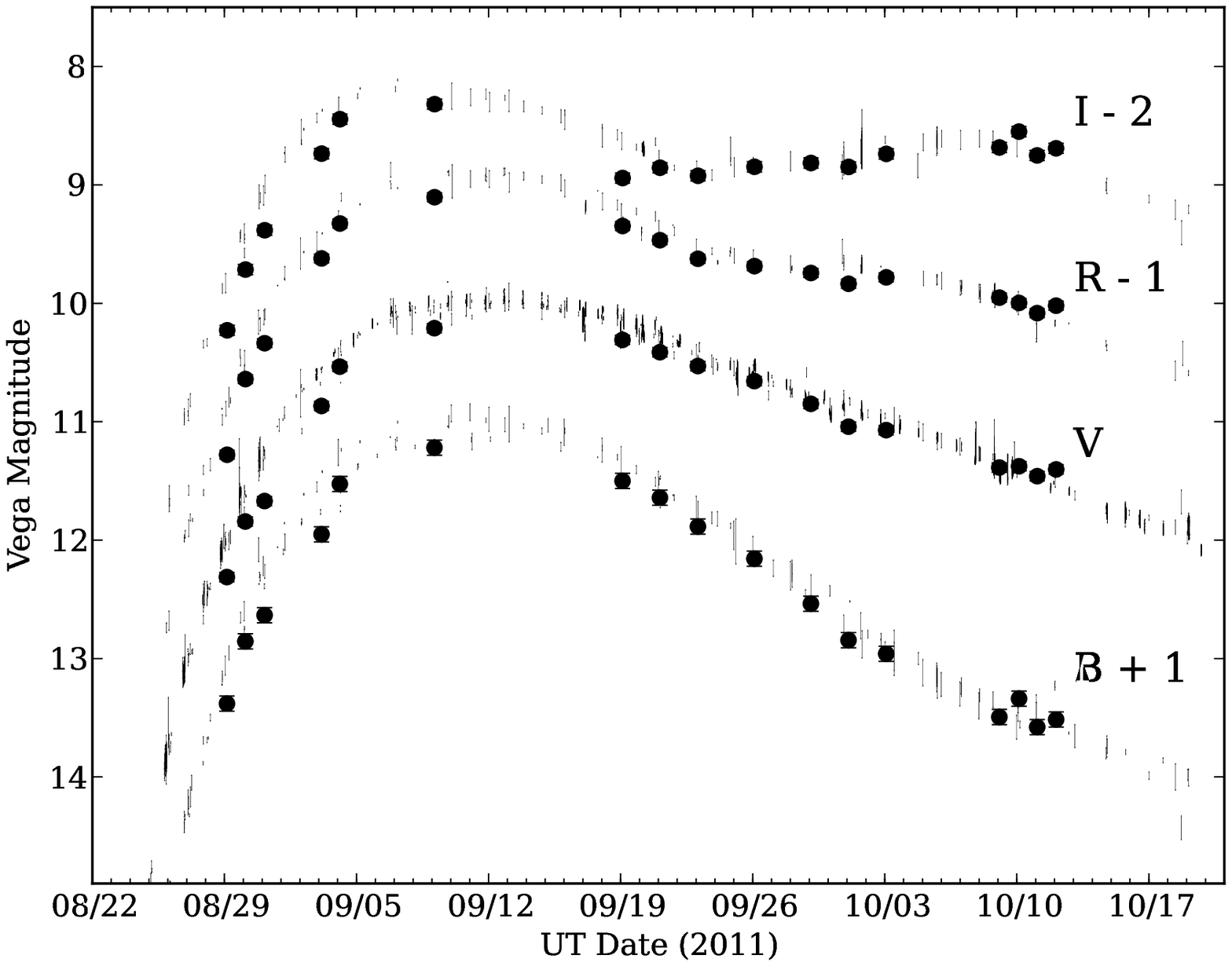}
	\caption{\footnotesize Preliminary early-time light curve of the SN~2011fe from RCT data (solid points) from~\cite{Gott:uq}, with updated calibration derived in this work. Data reported to AAVSO is also shown for comparison.}
	\label{fig:11fe}
\end{figure*}

The photometric calibration coefficients were determined with total errors in the mean of less than $0.3\%$.  The photometric errors (or errors in the calibration, presented as $1\sigma$ errors in Table~\ref{tab:coeffs2}) are also very reasonable, although there may be systematic contributions to these residuals attributable to faults in the facility: a likelihood further discussed in \S\ref{sec:caveats}.  As an independent test, these photometric corrections were applied to the early RCT direct imaging data of SN~2011fe, obtained at average airmass of 1.9. The results shown in Figure~\ref{fig:11fe} are in excellent agreement with data from various observatories of this nearby Type~Ia supernova.

\begin{figure*}[t]
	\centering\includegraphics[width=5.5in]{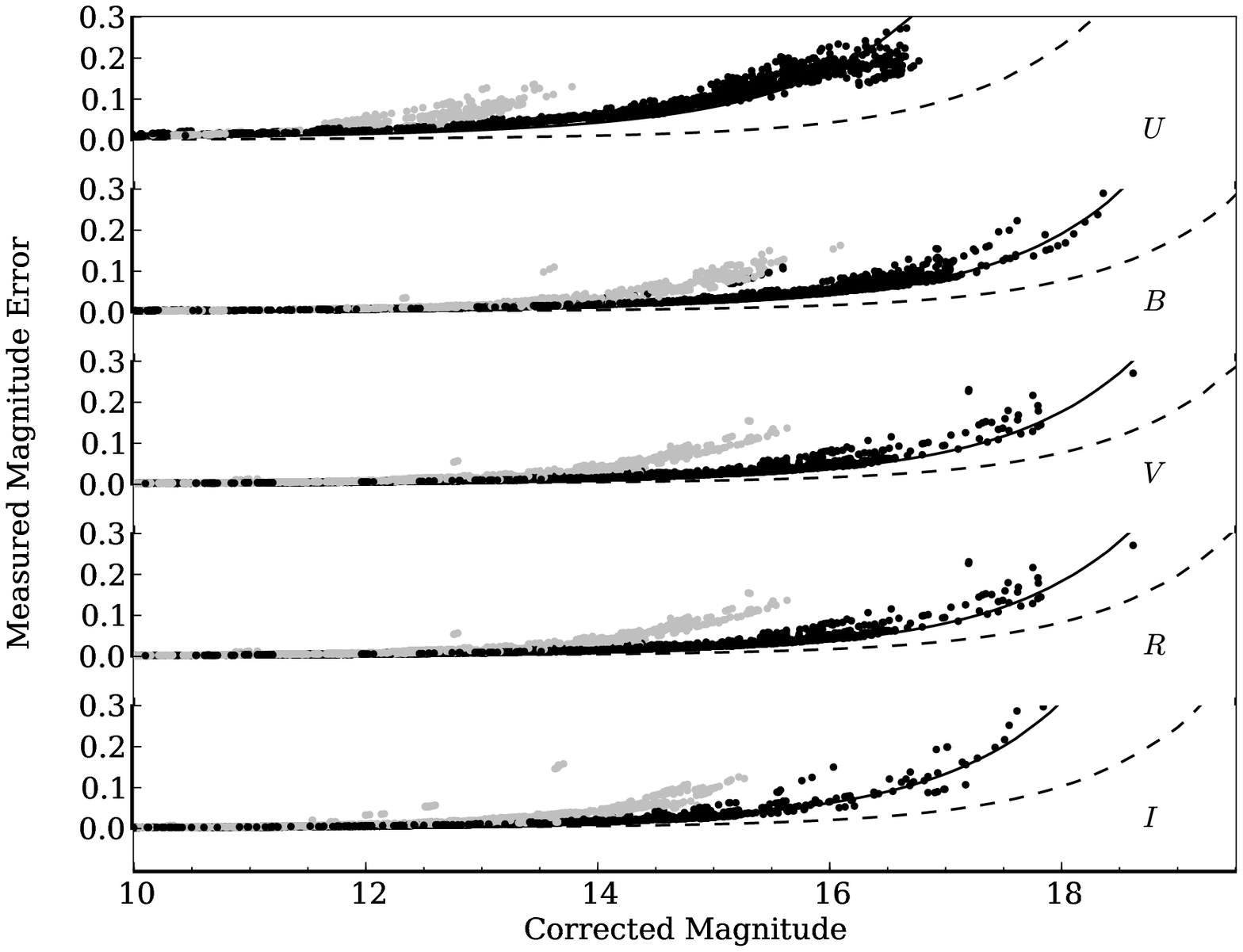}
	\caption{\footnotesize Measured magnitude errors a function of final calibrated magnitudes. The dashed lines are derived from signal-to-noise estimates from the RCT exposure time calculator, for exposures of 30~s, 20~s, 10~s, 10~s, and 10~s for each the {\it U, B, V, R,} and {\it I}-bands, respectively, under average sky brightness conditions. The solid lines are the same estimates, scaled by loss in S/N factors (see \S~\ref{sec:etc}). Gray points show high noise data taken between 2010 December 26 and 2011 December 31, and black points are all other data.\label{fig:snr}}
\end{figure*}
	
As another test of the calibration quality, we compared the calibrated magnitudes of all $22,059$ data points to their measured photometric errors, in a signal-to-noise comparison shown in Figure~\ref{fig:snr}. The figure shows an orderly agreement between the calibrated magnitudes and the measured magnitude errors that would otherwise be absent (resulting in increased scatter along the abscissa) if the calibration were inaccurate. There are two issues, however, to note. First, the trend in increasing noise occurs at much shallower magnitudes (by factors of $2-4$) than expected from estimates in signal-to-noise given by the RCT-ETC, given the appropriate exposure times, seeing conditions, and average sky brightnesses. Some deviation from these ideal estimates are expected, as the overall system response changes over time. It is not likely this is an issue with the aluminization of the primary, as measures of the reflectivity of the primary mirror have changed little since 2009, and are appropriately accounted for in the RCT-ETC models. The atmospheric corrections used in the RCT-ETC are also well matched to the {\tt MODTRAN5} models for the site. It is most likely this is a degradation in the broadband filter A/R coatings, reducing the response since their initial measures over a decade ago. 

The second issue to note is the spur of high-noise data seen at relatively bright magnitudes in each passband, indicated in gray in Figure~\ref{fig:snr}. These data points are correlated in time,  uniquely associated with a range of dates between 2010 December 26 and 2011 December 31. They do not consistently correspond to any range in photometric quality, either in $cloud$ value or residual in calibration, as is seen in Figure~\ref{fig:zp}. Maintenance logs over the same period indicate a herring-bone noise pattern in the biases that was later attributed to a failing (and eventually failed) power supply card in the camera controller, that was subsequently replaced. Measures of the read noise in that time frame were consistently an order of magnitude larger than at other more nominal times. These noisier data do not have a substantial effect on the determination photometric calibration coefficients. They do, however, add to the overall measurement error budget.

\subsection{Caveats in Photometric Calibration}~\label{sec:caveats}
Throughout this analysis there were concerns over possible errors or failures of the remote observatory, and the  degradation of the facility over the long refurbishment, as budgets limited servicing and regular maintenance schedules. This calibration program was implemented, in part, to monitor the working of the observatory, and the results indeed show larger intrinsic scatter than one might expect. 

The largest source of systematic error is likely attributable to the RCT dome. The RCT has the same dome that was installed in 1965 for the Remote Controlled Telescope, just wide enough to accommodate the full entrance pupil of the telescope. Slight errors in dome azimuth encoder positions partially occult the opening angle, and dim target sources. With few redundant monitors, a problem with the encoders went largely unnoticed until observations in late 2011 of a very bright target produced secondary, out-of-focus reflection images off the dewar window, which were clearly knife-edged. The problem was later identified as an intermittent failure of the PMAC card, specifically on the channel which controlled and monitored the dome encoder. The card has since been replaced. While it is reasonable that dome occultation could have been responsible for the fainter loci of residuals seen in Figure~\ref{fig:zp}, it remains unclear how many images were previously affected. It is possible, however, that the amount of occultation could been consistent at repeated azimuthal positions.
\begin{figure*}[t]
	\centering\includegraphics[width=5.5in]{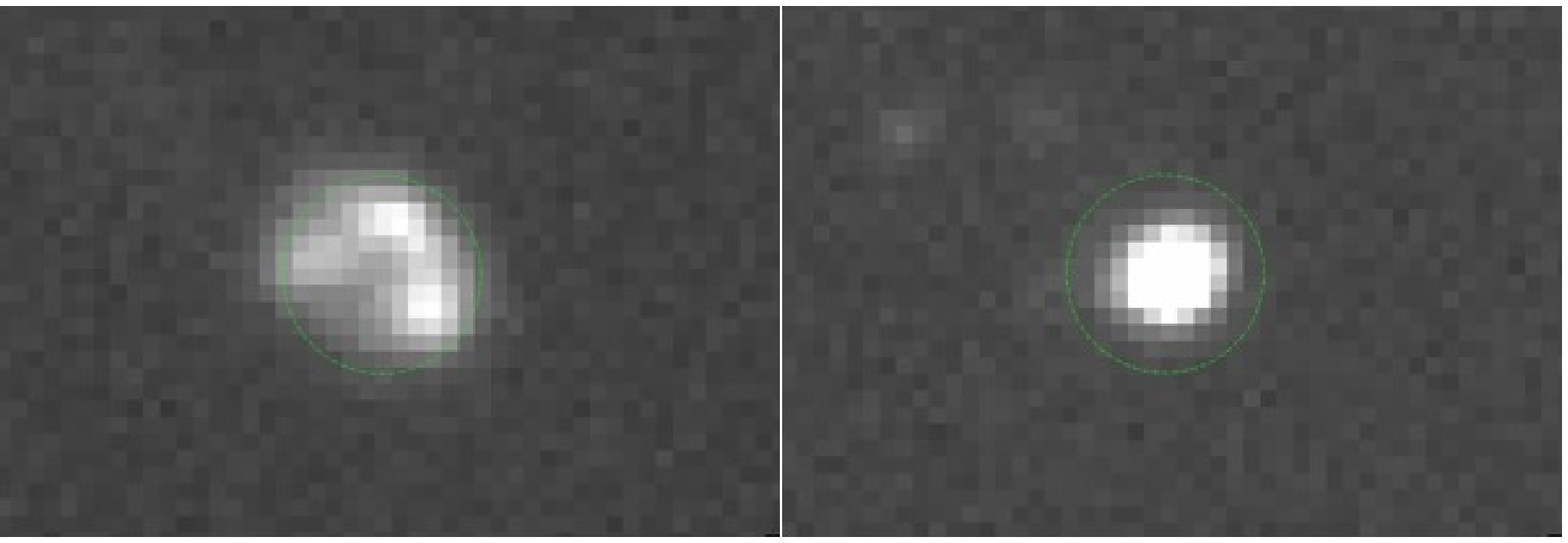}
	\caption{\footnotesize Comparison of the PSF under poor and good conditions. The images are of the same source taken when ({\it left}) both the focus was soft and the airbags supporting the primary mirror were completely deflated, and ({\it right}) under normal operating conditions. The poor image represents a worst-case scenario for photometry in this analysis, when more source light has overflowed the $3.\arcsec71$ aperture (shown in green) than in nominal conditions.}\label{fig:psf}
\end{figure*}

There are other systematic concerns, but at a much lower level. Two instrumental issues led to significant variations in PSF. The focus encoders have seasonal drifts that require manual assessment and correction, and at times this calibration was late. These few, slightly out of focus images were generally not a concern to this large aperture photometry. In addition, the mirror support cells were at times under-inflated, due to a problem with the air compressor and the bladder system that also contributed to an unusual stellar shape. A comparison of the image quality under these conditions to those under more nominal conditions is shown in Figure~\ref{fig:psf}. Clearly PSF photometry would not be possible with these complex issues in optical quality, which is why large aperture photometry was used in this analysis. Tests on the growth curves, exemplified in Figure~\ref{fig:growth}, show the selected apertures were large enough to enclose approximately 99\% of the source light, even with these variations in PSF. The standard star targets were not in crowded fields, and source blending was not an issue. The variations in PSF due to seeing at the site was also not a factor. The issue with the mirror support system has been addressed, but the focus still requires seasonal human intervention. 

Lastly, there are concerns over the loss of reflectivity (or additional scattering) due to dust settling on the telescope optics, and oxidation of the primary and secondary cells. Measures of reflectivity and scattering following a cleaning in 2009, and again in 2012, show as much as a $10$\% loss of signal over the same period, affecting bluer wavelengths more than red. Empirically, there does not appear to be an appreciable trend (to $>0.1$ mag) over the period from late 2011 to the end of 2012, when the majority of data for this analysis was obtained. There is a clear decline, however, from 2012 to 2013, and no data from that period is used in our final calibration analysis.

\section{Assessment Summary}

The RCT 1.3-meter is now fully operating and poised to continue its contribution to a diversity of science programs. As presented, the telescope meets most of the expectations of its initial science requirements, and has demonstrated reliability, with more than 4 years of continuous operation. It supports a wide suite of multipurpose observations, facilitated by its efficient automated scheduler and the comprehensive suit of filters continuously available.

The photometric quality of the facility is commensurate with other comparable imaging facilities at KPNO, with errors in coefficients of less than 0.3\%, and total error in the mean calibration of less than $10\%$. Although this photometric error is somewhat large, the independent comparison of calibrated RCT data to other measures on SN~2011fe imply a better precision than the total error in the mean calibration would seemingly indicate. There were identifiable instrumental errors that likely detracted from the measured precision, all of which have since been rectified. It is, therefore, expected that data obtained after 2013 will have improved photometric precision.

More work will be needed to verify that the available cloud monitor can be reliable for isolating photometric data in future work. In future calibrations, it may be useful to do an analysis of the photometric quality using gray extinction (due to clouds) as an additional free parameter in fields of high stellar density, with comparisons to data from the Sloan Digital Sky Survey or PanSTARRS. The Landolt standard fields selected for this analysis generally lack a sufficient stellar density within the RCT field of view to apply such an analysis to these observations. Additionally, considerable care will be needed in the color transformations from the Sloan filter set to the RCT broadband set~\citep[e.g.]{Jester:2005fk} to minimize this additional error component.

We thank our anonymous referee for very valuable comments and suggestions. We also thank the various staff at Brevin Technologies, LLC., KPNO, and NOAO for their contributions to the refurbishment, and continuing work servicing and maintaining the RCT facility. The refurbishment of the RCT was made possible by NASA grant NAG58762. DKW acknowledges support from the NSF PAARE and AST-0750814. The RCT 1.3-meter is operated by the RCT Consortium, and jointly run by Western Kentucky University, South Carolina State University, and Villanova University. IRAF is distributed by the National Optical Astronomy Observatories, which are operated by the Association of Universities for Research in Astronomy, Inc., under cooperative agreement with the National Science Foundation.

\bibliography{strolger}{}

\end{document}